\documentclass[aps,prd,twocolumn,superscriptaddress,showpacs]{revtex4}
\usepackage{graphicx}
\usepackage{dcolumn}
\usepackage{bm}
\usepackage{natbib}
\usepackage{multirow}

\newcommand{\beq}{\begin{equation}}
\newcommand{\eeq}{\end{equation}}
\newcommand{\beqa}{\begin{eqnarray}}
\newcommand{\eeqa}{\end{eqnarray}}

\newcommand{\lyaf}{Ly$\alpha$ forest}

\def\affilmrk#1{$^{#1}$}
\def\affilmk#1#2{$^{#1}$#2;}

\def\ptonp{1}
\def\ictp{2}
\def\cita{3}
\def\pton{4}

\begin{document}

\title{Can sterile neutrinos be the dark matter?}  
\author{
Uro\v s Seljak \affilmrk{\ptonp,\ictp},
Alexey Makarov \affilmrk{\ptonp},
Patrick McDonald \affilmrk{\cita},
Hy Trac \affilmrk{\pton,}
}
\address{
\parshape 1 -1cm 20cm
\affilmk{}{Physics Department, Princeton University, Princeton, NJ 08544,
USA}
\affilmk{\ictp}{International Center for Theoretical Physics, Trieste, Italy}
\affilmk{\cita}{Canadian Institute for Theoretical Astrophysics, University of Toronto, ON M5S 3H8, Canada}
\affilmk{\pton}{Princeton University Observatory, Princeton, NJ 08544,USA}
}


\date{\today}

\begin{abstract}
We use the Ly-$\alpha$ forest power spectrum measured 
by the Sloan Digital Sky Survey (SDSS) and high-resolution 
spectroscopy observations 
in combination with cosmic microwave background
and galaxy clustering constraints to place limits on a 
sterile neutrino as a dark matter candidate in the 
warm dark matter (WDM) scenario. 
Such a neutrino would be created in the early universe 
through mixing with an active neutrino and would 
suppress structure on scales smaller than its free streaming scale. 
We ran a series of high-resolution hydrodynamic simulations 
with varying neutrino mass to 
describe the effect of a sterile neutrino on the Ly-$\alpha$ forest 
power spectrum. 
We find that the mass limit is $m_s >14 {\rm keV}$ at 95\% c.l.
(10keV at 99.9\%), which is nearly an order of magnitude tighter constraint 
than previously published limits and is
above the upper limit allowed 
by X-ray constraints, excluding this candidate as dark matter in this model.
The corresponding limit for a neutrino that decoupled 
early while in thermal equilibrium
is 2.5keV (95 \% c.l.). 

\end{abstract}


\maketitle

\setcounter{footnote}{0}


One of the major unsolved mysteries in cosmology is the nature of the
dark matter in the universe. Observational evidence points towards 
cold dark matter (CDM), for which random velocities are negligible. 
Two of the leading particle physics 
candidates, the lightest supersymmetric partner 
and axions, both require extensions beyond the standard model. 
At the same time, neutrino experiments over the past decade 
have shown that neutrinos oscillate from one flavor to another, 
which is only possible if they have mass.  
Current data from atmospheric and solar neutrino experiments 
\cite{2004PhRvL..93j1801A,2001PhRvL..87g1301A}
are compatible with mixing between the three active neutrino 
families. Perhaps the simplest way
to incorporate these neutrino phenomena into the 
standard model is to add right-handed neutrinos, just as for other 
fermions. 

Given this extension of the standard model it is natural to ask if 
these (almost) sterile right-handed neutrinos can also explain the dark matter 
\cite{1994PhRvL..72...17D}. 
At least two sterile neutrinos are required 
to explain the origin of neutrino mass and existence of 
different mass mixing scales in solar and atmospheric 
neutrinos, so in a model with three families of sterile neutrinos 
a third one can act as dark matter \cite{Asaka:2005an}. 
Such neutrinos 
free stream and erase all fluctuations on scales smaller 
than the free streaming length. This length is roughly 
proportional to the temperature and inversely proportional to the mass 
of neutrinos.
Thus if the neutrino mass is sufficiently high, or the temperature 
sufficiently low, then it acts just like CDM 
and can satisfy all of the observational constraints from 
structure formation. Current constraints require the neutrino 
mass to be above 1.8keV 
\cite{2005PhRvD..71f3534V,2005astro.ph.12631A}. This is below the 
5-8keV upper limits 
from the absence of detection of X-ray photons from
radiative decays \cite{2001ApJ...562..593A,2005astro.ph.11630A,2005MNRAS.364....2M,2006hep.ph....1098B}. 
A massive
neutrino in the keV range has also been suggested as a possible explanation for 
high pulsar velocities 
\cite{1997PhLB..396..197K} and 
such a model can possibly
explain baryon asymmetry in the universe 
\cite{Asaka:2005pn}. 

A sterile neutrino is not completely sterile if it is 
to provide the origin of mass for active neutrinos: it interacts with 
active neutrinos and the interaction strength is parametrized
by the active-sterile mixing angle $\Theta$, which in this model is 
required to be very small, $\Theta<10^{-4}$. In this regime 
sterile neutrinos never reach thermal equilibrium \cite{1994PhRvL..72...17D}. 
In general a sterile neutrino decays into active ones, but the lifetime 
can be well above the age of the universe over a broad range of masses 
and mixing angles of interest, so it is effectively stable. 
If the interaction rate is energy independent then 
the momentum distribution  
of sterile neutrinos is simply a reduced version of the 
distribution of active neutrinos 
\cite{1994PhRvL..72...17D}.
In practice the interaction rate is not constant over the range of 
masses of interest, because at temperatures 
above the QCD transition more interaction channels become available
\cite{2002APh....16..339D,2001PhRvD..64b3501A,2005astro.ph.11630A}. 
In this paper we use the 
latest calculation \cite{2005astro.ph.11630A}, which  
however 
has only a minor effect relative to the constant interaction rate, 
reducing the derived mass limits by 
about 10\% 
\footnote{
We assume throughout the paper 
a negligible lepton asymmetry and do not consider the 
possibility of a resonant enhancement in neutrino production, nor do we
consider the possibility for additional creation processes of
the sterile neutrino, such as the coupling to inflaton.}.

For keV masses of interest the corresponding 
free-streaming length is of order a Megaparsec (Mpc) and below. 
Distinguishing between cold and warm dark matter thus requires a sensitive 
probe of linear fluctuations on small scales, but 
nonlinear evolution erases the initial 
conditions on these scales today. 
Of the current tracers 
of density fluctuations the one that is most suitable for WDM is 
the Ly-$\alpha$ forest \cite{2000ApJ...543L.103N}. It is 
measured from the absorption observed in quasar spectra
by neutral hydrogen in the intergalactic medium 
and has been shown to accurately trace the dark matter distribution 
\cite{1998ApJ...495...44C}.
It probes fluctuations down to sub-Mpc scales at redshifts between
2 and 4, so nonlinear evolution, while not negligible, has not erased
all of the primordial information.
Current WDM constraints from the \lyaf\ \cite{2005PhRvD..71f3534V}
do not include the latest measurements of the \lyaf\ from the
Sloan Digital Sky Survey (SDSS) \cite{2004astro.ph..5013M} \footnote{The 
exception 
is \cite{2005astro.ph.12631A}, where however the linear power spectrum 
determination of \cite{2005ApJ...635..761M}, only valid for
CDM, was applied to WDM models, making the results
unreliable.}. 
The goal of this letter is  to derive new limits by
incorporating these observational constraints and combining them with 
a series of new hydrodynamic simulations which accurately describe the 
effect of a massive neutrino on the Ly-$\alpha$ forest. 



\begin{figure}
\includegraphics[width=3in]{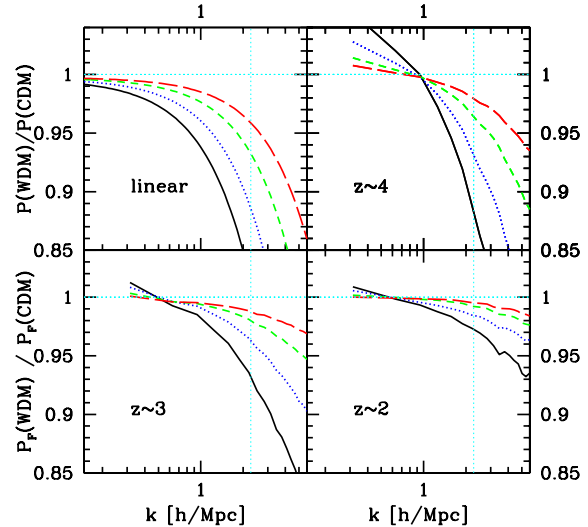}
\caption{
Ratio of WDM power spectrum relative to CDM shown over the relevant 
observational range. 
From left to right the sterile neutrino masses are 
6.5keV, 10keV, 14keV and 20keV. Top left corner shows 
1d linear power spectrum, while the other 3 panels 
show the ratios from hydrodynamic simulations at redshifts 
2, 3 and 4. 
We used concordance cosmology with 
$\Omega_{m}=0.28$ and $H_0=71{\rm km/s/Mpc}$. Dashed 
vertical line shows the upper limit on $k$ for SDSS data.} 
\label{fig1}
\end{figure}

The linear theory calculations of WDM using 
CMBFAST \cite{1996ApJ...469..437S} 
result in the matter power spectra shown, relative to CDM,
in the upper left panel of figure \ref{fig1}. 
We plot the ratio of WDM to CDM power  
for $m_{\nu}=6.5$, 10, 14 and 20keV. 
Instead of the usual 3d power spectrum we 
plot the corresponding 1d projection, which is more relevant for the 
comparison to the 1d \lyaf\ observations. 
One can see the suppression of power on scales smaller than the 
free-streaming length, which depends on the neutrino mass. 
While for $m_{\nu}>10$keV 
there is hardly any effect for $k<5$h/Mpc
in 3d (see e.g. relevant 
figures in \cite{2005PhRvD..71f3534V,2005astro.ph.12631A}), the corresponding 1d power spectrum 
shows more of an effect because small scale modes in 3d are projected
to large scale modes in 1d. For example, 
for $m_{\nu}=20$keV there is essentially no effect in 3d for $k<3{\rm h/Mpc}$
and even at $k=5{\rm h/Mpc}$ the power suppression is only 2\%. 
SDSS measurements of the flux power in 1d do not extend above 2h/Mpc and high 
resolution spectra are reliable up to 5h/Mpc. 
So if one were interpreting them as measuring 3d power 
then it would be very difficult to detect neutrino masses 
in this mass range. However, the corresponding 1d case in figure \ref{fig1} 
shows a 3\% power suppression 
at 2h/Mpc and 15\% at 5h/Mpc. This rapidly increases with declining mass, so 
that for a 6.5keV neutrino the 
power suppression is 15\% at k=2h/Mpc and a factor of 2 at k=5h/Mpc. 

Nonlinear evolution and hydrodynamic effects further modify the 
linear predictions, which must be addressed with 
simulations. 
We ran hydrodynamic simulations for a series of neutrino masses 
ranging from 3.4keV to 20keV.
Many convergence tests and 
comparisons between different hydrodynamic codes have been 
performed, which will be presented in a separate publication. 
These tests confirm the accuracy of the original analysis in  
\cite{2005ApJ...635..761M}, which 
was based on a grid of hydro-PM simulations sparsely 
calibrated with hydrodynamic simulations. 
For the hydrodynamic simulations in this paper we used the 
Eulerian moving frame TVD+PM code described in \citep{2004NewA....9..443T}.  
The Eulerian conservation equations are solved in a frame moving with the fluid where numerical Mach numbers are minimized, allowing thermodynamic variables to be accurately calculated for both subsonic and supersonic gas.
Our standard simulations used 20 Mpc/h boxes with $256^3$ particles for dark
matter and $512^3$ cells for gas.  We used 10 Mpc/h boxes with equal or twice
this resolution to test convergence.

Simulation results are shown in figure \ref{fig1} 
for redshifts 2, 3 and 4 that span the observational range. 
We have adjusted the level 
of the UV background to match the mean absorption as measured from the data. 
The results show that for $m_{\nu}=20$keV there are 1-2 percent effects 
at k=2h/Mpc at z=4, increasing 
to 11\% for $m_{\nu}=6.5$keV. 
At k=5h/Mpc
the effects are 6\% 
suppression for $m_{\nu}=$20keV mass and  
a factor of 1.5 for $m_{\nu}=$6.5keV. 
These are redshift dependent: while 
there is little differentiating power between models
at low redshift, the differences become significantly larger 
at high redshift, where the mean level of absorption is higher and the 
linear power is better preserved.  
Finally, we note that the suppression of 
small scale power also affects the large scale 
bias of the flux power spectrum, 
which explains why the ratios do not converge to unity on large scales.

Since the free-streaming scale for these high mass neutrinos is so small 
one must be careful to have sufficient spatial resolution
to capture the suppression of power on small scales. This is not a 
trivial requirement: for a 10keV neutrino the scale corresponding to 
a 50\% suppression of the WDM transfer function relative to CDM occurs 
at 4 times the cell size in our standard simulations
(1/4 of the Nyquist frequency -- for a 20keV neutrino the 
corresponding number is 1/2).  Doubling the  
resolution does not change our results significantly ($<20$\% of 
the size of the WDM effect), so our simulations have 
converged over the relevant mass range, but only barely. 
It is important to note that insufficient resolution 
weakens the constraints, since the power suppression is 
not captured on scales below the Nyquist scale corresponding to the cell size. 
This could be an issue for the SPH simulations used in 
\cite{2005PhRvD..71f3534V}: for 
example, for their 10keV sterile neutrino mass simulation (corresponding to
2keV thermal neutrino in their plots) they find essentially no effect in
the \lyaf\ on the observable scales, 
concluding that such masses cannot be probed by the \lyaf, in clear 
contradiction with our results in figure \ref{fig1}. 
In their simulations a 50\% suppression of the 10 keV WDM transfer function 
relative to CDM happens at the scale of the particle separation.  This is 
a factor of 4 worse resolution than our cell size for gas, and a factor of 2
worse than our particle spacing for dark matter.  
Thus much of the power suppression is missing already in the initial 
conditions and the Lagrangian spatial resolution in the SPH simulations cannot 
restore it (additionally, it is unclear that the Lagrangian nature of the SPH 
should help for the near-mean or even underdense gas probed by 
the \lyaf\ at $z\sim 4$). 


In addition to the Ly-$\alpha$ forest flux 
power spectrum from SDSS
\cite{2004astro.ph..5013M}
we have added earlier high-resolution \lyaf\
constraints 
in a weak form 
\cite{2000ApJ...543....1M,2001ApJ...562...52M}.
When testing the robustness of the derived constraints 
we also include the more recent high-resolution \lyaf\ data
\cite{2002ApJ...581...20C,2003astro.ph..8103K}. 
While galaxy clustering and CMB data do not constrain WDM, they are useful 
for constraining the remaining cosmological parameters. 
We use as inputs the SDSS galaxy power spectrum \cite{2004ApJ...606..702T}
and CMB power spectrum from WMAP 
\citep{2003ApJS..148..135H}.
Our analysis is based on 
the Monte Carlo Markov Chain (MCMC) method
\citep{2003MNRAS.342L..79S} 
and uses
CMBFAST
\cite{1996ApJ...469..437S}
to output
both CMB spectra and the corresponding matter power spectra $P(k)$.
The output transfer
functions  
are interpolated onto a grid of simulations 
using the matter power spectra rather than the 
neutrino mass, since it is the matter spectrum that is most directly related 
to the observations. 

Our most general cosmological parameter space has 9 parameters, 
which are the Hubble constant, matter and baryon density, amplitude, 
slope and running of the primordial power spectrum, tensor to scalar 
ratio, optical depth and neutrino mass. 
Since in most models of inflation tensors
and running are expected to be small we also explore the constraints 
when they are set to zero.  

We compare the theoretical $P_F(k)$
directly to the measured power spectrum. This is particularly important for 
the WDM analysis, where 
one cannot use the 3d linear power spectrum amplitude and slope
constraints
as given in \cite{2005ApJ...635..761M}, since as emphasized there 
these are valid only in the 
context of standard CDM models without WDM.  
The \lyaf\ contains several nuisance parameters which we are not interested
in for the cosmological analysis, so they are marginalized over. 
These include the UV background intensity,  
temperature-density relation of the gas and the filtering length (related to 
Jeans scale \citep{1998MNRAS.296...44G}). 
We also include a marginalization over several additional 
physical effects, such as fluctuations in the UV background and galactic 
winds \cite{2005ApJ...635..761M,2005MNRAS.360.1471M}.

\begin{figure}
\includegraphics[width=3in]{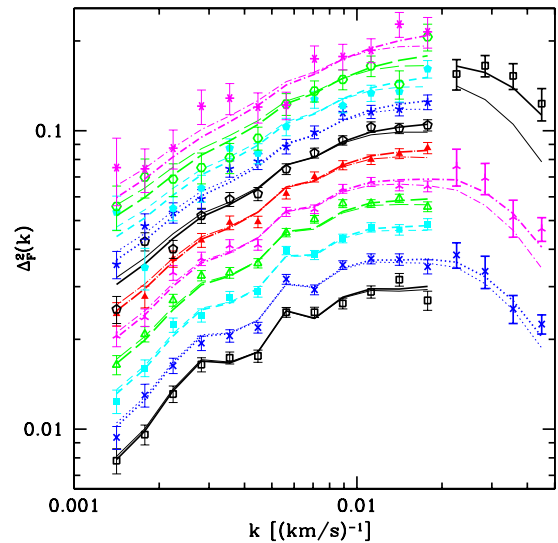}
\caption{To the left are the observed SDSS
\lyaf\ flux power spectra as a function of redshift from 
2.2 (bottom) to 4.2 (top) in steps of 0.2. To the right are the 
power spectra from the 
high resolution data compiled at redshifts 2.4, 3.0 and 3.9. 
For each redshift 
the thick lines are from the best fitted CDM model, while the 
(generally lower at high k) thin lines are for the corresponding 
WDM model with 6.5keV 
sterile neutrino. The latter is discrepant with 
both SDSS and high resolution data, 
with most 
of the distinguishing power coming from higher redshifts. 
} 
\label{fig2}
\end{figure}

Applying the standard MCMC analysis to the WDM case we find 
no evidence of WDM: the limit is
$m_{\nu} > 14{\rm keV}$ at 95 \% c.l. (10keV at 99.9\% c.l.). 
The corresponding limit for  neutrinos which were in 
thermal equilibrium at a high temperature, when the universe had more 
degrees of freedom, and then decoupled, is
$m_{\nu}<2.5$keV
at 95\% c.l.
This constraint is obtained in our 9 parameter space, but  
reducing the parameter space to the minimal 7 parameters
without running and tensors does not change the results. 
The \lyaf\ data, best fitted CDM model and corresponding WDM model 
for $m_{\nu}=6.5$keV are shown in figure \ref{fig2}. 
One can see how the suppression of power on small scales 
in WDM makes the fit worse.  For this figure, where we have not adjusted all 
the other parameters to their best fitted value, 
the increase in $\chi^2$ with
WDM is 77 -- when the \lyaf\ model parameters and power spectrum amplitude 
and slope are fitted, $\Delta \chi^2$ is still 27.  
Even without high resolution constraints, the poor 
fit to the SDSS data is apparent, especially at higher redshifts.  
Removing the high resolution data only weakens the bounds by 15\%. The 
converse however is not true: without SDSS the previously found constraint
(after 10\% adjustment for nonthermal momentum 
distribution) is $m_{\nu}>1.8$keV (95\% c.l.) \cite{2005PhRvD..71f3534V}.
This is because 
within the high resolution
data 
there are degeneracies between WDM and 
many of the nuisance 
parameters such as the temperature of the IGM, UV flux and filtering scale. 
These can be removed by adding the large scale
flux power spectrum measured by the SDSS data. 
Finally, we note that 
using 
the more recent high-resolution \lyaf\ data\cite{2002ApJ...581...20C,2003astro.ph..8103K} 
does not improve the limits obtained above.

Sterile neutrinos that couple to active ones also decay 
and their radiative decays result in photons with energy peaking at close to 
one half of the neutrino mass, which for keV masses can be searched for in 
X-rays
from either clusters or from their cumulative contribution 
in a random direction. 
Absence of such X-ray emission in the Virgo cluster
results in an upper limit on the mass of 8keV \cite{2005astro.ph.11630A}
while recent reevaluation of X-ray background constraints 
gives an upper limit of
5keV for the value of mixing angle that 
matches the required density of sterile neutrinos
\cite{2006hep.ph....1098B}. 
These are all below our 99.9\% 
lower limit, suggesting sterile neutrinos cannot be the dark 
matter in this model. 

Can the bounds presented here be invalidated by some 
additional physical effect in \lyaf\ that is not included in our model? 
This is unlikely, but cannot be ruled out completely. 
 There are possible physical effects that can in principle affect the \lyaf\
power spectrum and while most of them have been shown to be negligible or 
are already part of our standard analysis \cite{2005MNRAS.360.1471M,2005astro.ph.10841L},
there remains a possibility 
that something else will turn out to be 
important. 
However, it is 
important to recognize how successful is the current model
in explaining the observations. Any 
potential effects that may be missing in the current 
analysis are constrained by the remarkable agreement of the 
simplest CDM model with the data. 
It seems unlikely that if we lived in a WDM universe its signature 
were erased exactly by some (yet to be discovered) physical effect.
Barring any such cancellations we may conclude that the simplest
model of sterile neutrinos as 
the dark matter 
is ruled out, since the upper limit from their decays
and the lower limit from their effect on large scale structure
no longer leaves an open window. 

U.S. is supported by the
Packard Foundation, 
NASA NAG5-1993 and NSF CAREER-0132953.  
H.T. acknowledges NASA grant NNG04GC50G.
We thank Alexei Smirnov for useful discussions. 

\bibliography{cosmo,cosmo_preprints}
\end{document}